\documentclass[12pt,a4paper,a4paper,a4paper,a4paper,twoside,a4paper,a4paper]{iopart}
\usepackage[T1]{fontenc}
\usepackage[utf8]{inputenc}
\usepackage{nicefrac}
\usepackage{graphicx}
\usepackage{amssymb}

\makeatletter

\usepackage{array}

\usepackage{nicefrac}

\makeatletter

\usepackage{array}

\usepackage{nicefrac}

\makeatletter

\usepackage{nicefrac}

\makeatletter

\usepackage{nicefrac}

\makeatletter
\usepackage{nicefrac}

\usepackage{textcomp}

\makeatletter

\usepackage{iopams}

\usepackage{cite}

\usepackage{eurosym}

\usepackage{multicol}

\usepackage{cropmark}

\makeindex

\newcommand{\be}{\begin{equation}} \newcommand{\ee}{\end{equation}}
\newcommand{\beas}{\begin{eqnarray*}}
\newcommand{\eeas}{\end{eqnarray*}}
\newcommand{\bea}{\begin{eqnarray}}
\newcommand{\eea}{\end{eqnarray}}

\usepackage{dcolumn}

\usepackage{bm}

\makeatother

\makeatother

\makeatother

\makeatother

\makeatother

\makeatother

\begin{document}

\title[Opinion Dynamics with Learning]{Opinion Dynamics of Learning Agents: Does Seeking Consensus Lead
to Disagreement? }

\author{Renato Vicente\textonesuperior{}, André C.R. Martins\textonesuperior{}
and Nestor Caticha\texttwosuperior{} }

\address{\textonesuperior{} GRIFE-EACH, Universidade de São Paulo, Campus
Leste, 03828-080, São Paulo-SP, Brazil}

\address{\texttwosuperior{} Dep. de Física Geral, Instituto de Física, Universidade
de São Paulo, Caixa Postal 66318, 05315-970, São Paulo-SP, Brazil}

\ead{rvicente@usp.br}

\begin{abstract}
We study opinion dynamics in a population of interacting adaptive
agents voting on a set of issues represented by vectors. We consider
agents which can classify issues into one of two categories and can
arrive at their opinions using an adaptive algorithm. Adaptation comes
from learning and the information for the learning process comes from
interacting with other neighboring agents and trying to change the
internal state in order to concur with their opinions. The change
in the internal state is driven by the information contained in the
issue and in the opinion of the other agent. We present results in
a simple yet rich context where each agent uses a Boolean Perceptron
to state its opinion. If the update occurs with asynchronously exchanged
information among pairs of agents, then the typical case, if the number
of issues is kept small, is the evolution into a society thorn by
the emergence of factions with extreme opposite beliefs. This occurs
even when seeking consensus with agents with opposite opinions. If
the number of issues is large, the dynamics becomes trapped, the society
does not evolve into factions and a distribution of moderate opinions
is observed. The synchronous case is technically simpler and is studied
by formulating the problem in terms of differential equations that
describe the evolution of order parameters that measure the consensus
between pairs of agents. We show that for a large number of issues
and unidirectional information flow, global consensus is a fixed point,
however, the approach to this consensus is glassy for large societies. 
\end{abstract}

\newpage{}

\tableofcontents{}

\section{\label{sec:sec1} Introduction}

The ability of sustaining disagreement is a vital feature of a deliberative
democracy \cite{Johnson04}. Experiments show that humans seek conformity
to the opinions of those they interact (see \cite{Sunstein05} and
references therein) and observational studies of elections show that
disagreement is often sustained within close communication networks
of individuals~\cite{Huckfeldt95}. Despite strong micro-conformity
effects, democratic societies tend to sustain long term opinion diversity,
generating a micro-conformity versus disagreement puzzle.

The recent literature on opinion dynamics has mainly focused on mechanisms
for attaining consensus. Several attempts to model if consensus will
emerge in a model society have been proposed ~\cite{Castellano07,Galam82}.
Discrete models, like the Voter~\cite{Clifford73,Holley75} or Sznajd's~\cite{Sznajd00}
models always lead to consensus, at least in the long run and, in
that sense, are not good choices to describe disagreement. Social
Impact models~\cite{Latane81,Lewenstein92} allow for the survival
of diverse opinions, given a proper choice of parameters. While still
describing opinions about only one issue, in the Continuous Opinions
and Discrete Actions model (CODA)~\cite{Martins08a}, consensus is
only obtained at the local level, with the survival of divergent extreme
opinions in the society as a whole. This is achieved by distinguishing
between the agent internal probabilistic continuous opinion variable,
and what other agents observe, a discrete action. The internal probability
is updated by Bayesian rules and results from a history of previous
interactions between agents.

One important way in which the real world is different from these
models is that real people debate several issues simultaneously. In
the Axelrod Culture Model (ACM)~\cite{Axelrod97}, a set of issues
(culture) is represented by a vector of opinions. Agents influence
their neighbors on a lattice if there is agreement on a minimum number
of issues. The ACM has two absorbing states, one corresponding to
complete consensus and other to global disagreement, and a non-equilibrium
phase transition is observed having the degree of initial heterogeneity
as a control parameter \cite{Castellano00}. A modification of the
ACM proposed in \cite{Johnson04} is capable of sustaining a certain
level of local disagreement by introducing an auto-regressive influence,
with agents keeping track of each interaction and only being influenced
if there is sufficient evidence supporting the opinion being communicated.

Here, we propose to study patterns of disagreement in a population
of agents that adapt their internal representations to opinions on
a set of issues observed in a social neighborhood. Each agent has
its own adaptive decision mechanism. In particular we consider each
agent endowed of a neural network and a learning algorithm, which
is used to decide on public issues. The theoretical scenario we have
put forward is not limited by the nature of the information processing
units under consideration, nor by the underlying geometry of the graph
of social interactions. We could well study the interaction of probabilistic
units such as, for example, hidden markov models or of deterministic
neural networks. With the aim of advancing beyond general propositions
and making sharp statements, it is natural to limit its scope by considering
specific architectures. Therefore, in this paper we limit ourselves
to model each agent as a Simple Boolean Perceptron \cite{Engel01}
(an alternative involving Hopfield networks has been proposed in \cite{Stauffer08}),
a sufficiently simple machine which affords, on one hand, the obtention
of both analytical and numerical results. On the other hand it is
complex to the point of rich and interesting behavior.

Another important conceptual change with respect to other models is
that issues are represented by vectors. This allows to encompass complex
issues that cannot be pinpointed by a single number but maybe by a
set of numbers, each quantifying a single feature. An issue, in this
paper, is a point in a $N$ dimensional space. Actually we find it
more difficult to defend the modeling of an issue by a single number
than by a set. Consider real issues, e.g. loop gravity or strings,
government control over the stock market, pregnancy termination rights,
etc. Can they be represented by a single number each? More likely
different aspects, such as political, cultural, religious, historical,
ethical, economic, etc. should be taken into account while modeling
any of them. We however do not claim that can model such complex issues
by vectors, but rather make the claim that they certainly cannot be
modeled by a single number. A vector is the next natural step. We
also should notice that since agents debate more than an issue at
a time, our model can be regarded as one of cultural dynamics, akin
to the ACM. Our model is, however, distinct from ACM in at least one
very important way, namely, revealed opinions on diverse issues are
the result of the application of agent's hidden classification rule
that is modified upon social interactions, changing or reinforcing
these opinions.

In general terms we find that consensus is rarely attained even if
agreement is strongly favored in each interaction, that disagreement
is sustained locally in a probabilistic sense and when agents vote
on several issues simultaneously. In the less interesting case of
synchronous dynamics we find that disagreement is the norm, with consensus
only emerging in some particular regimes.

This paper is organized as follows. The next section provides a general
presentation of the model we propose. In Section 3 we analyze three
simple scenarios on a social graph defined by a one dimensional lattice
with periodic boundary conditions (ring). We present our conclusions
in Section 4.

\section{\label{sec:sec2} Opinion Dynamics in a Network of Perceptrons}

The model consists of $K$ interacting agents. The interaction is
defined by the fact that every agent can adapt its opinion formation
unit by learning from the opinion of other agents. We now describe
how an agent reaches its opinion on a given issue using its particular
neural network.

Perceptrons are usually studied within scenarios that include learning
a rule from a given data set (\emph{supervised learning)} or extracting
features from a given set of inputs (\emph{unsupervised learning)}
\cite{Engel01}\emph{.} A mutual learning scenario, where Perceptrons
try to learn from each other, was first studied in \cite{Metzler00}.
We here extend this work to the context of opinion dynamics. Our main
idea is that conformity related information processing may be modeled
by adaptive agents trying to predict their social neighbor opinion
$\sigma$ on a (multidimensional) public issue $\boldsymbol{x}$ drawn
from a set $\mathcal{X}$ of $P$ issues. A model of this sort simultaneously
allows for a number of features that are absent or only partially
present in models currently described in the literature, to say, it
allows for a dichotomy between internal representations (or beliefs)
and revealed opinions, for the introduction of memory effects, for
the definition of agreement in a probabilistic sense and for the attribution
of intensities to beliefs.

\begin{figure}[h]
\hspace{2.0cm}\includegraphics[clip,scale=0.36]{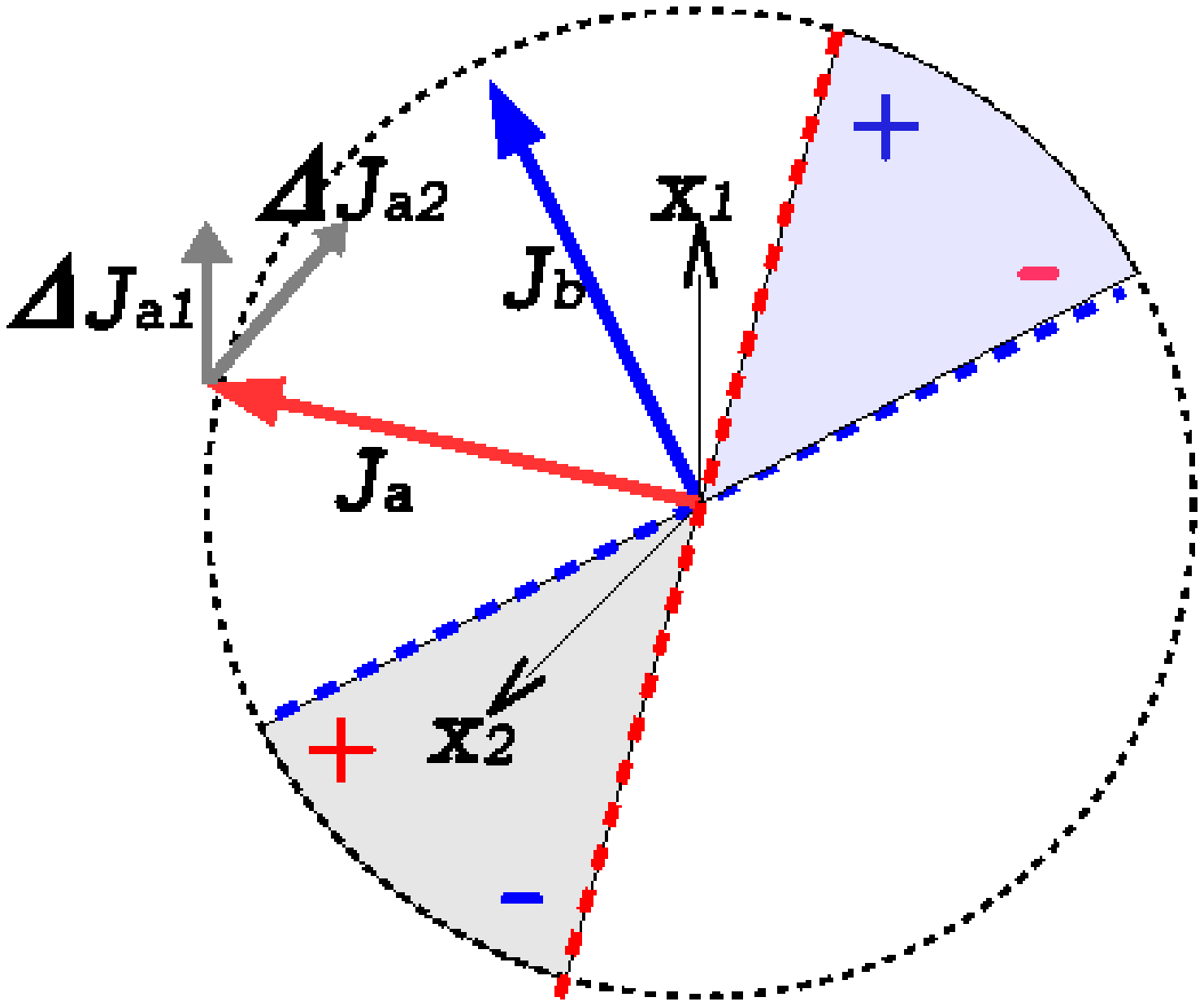}\hspace{0.6cm}\includegraphics[scale=0.65]{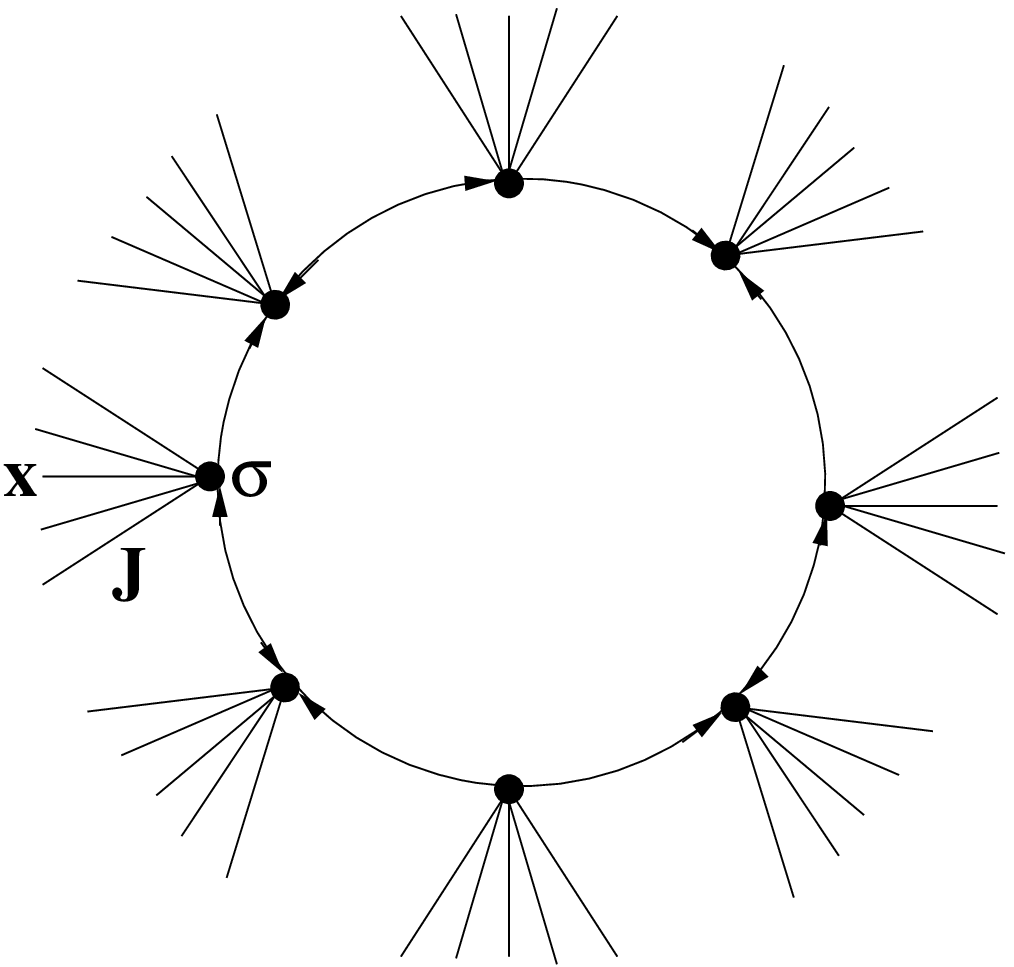}

\caption{\emph{(Left) Perceptron hyperplane and learning.} A Perceptron classifies
vectors ($\boldsymbol{x_{1}}$ and $\boldsymbol{x_{2}}$) in space
by defining hyperplanes (thick dashed lines) normal to the internal
representation vectors. In the figure two Perceptrons ($\boldsymbol{J}_{a}$
and $\boldsymbol{J}_{b}$) are shown, vectors with positive (negative)
overlaps are classified as $+1$ ($-1).$ In the shaded area Perceptron
classifications disagree. Vectors $\Delta\boldsymbol{J}_{a1}$ and
$\Delta\boldsymbol{J}_{a2}$ represent modifications of vector $\boldsymbol{J}_{a}$
upon the observation that agent $b$ agrees on classifying $\boldsymbol{x_{1}}$
as $+1$ but disagrees on the classification of $\boldsymbol{x_{2}}$.
Modifying vectors are, respectively, parallel and anti-parallel to
the vectors being classified with lengths proportional to the learning
rate $\eta$ times the modulation function \emph{$\mathcal{W}_{\delta}$.
(Right) Perceptrons on a ring.} The scenario we study places agents
on a digraph of social influences, such that $a$ is influenced by
$b$ if $b$ is a parent of $a$. In the figure each agent is represented
by a Perceptron (black circle) that acquires information on a public
issue $\boldsymbol{x}$, classifies it following an internal representation
vector $\boldsymbol{J}$ and reveals an opinion $\sigma$. Influence
can be unidirectional or bidirectional, synchronous or asynchronous.
\label{fig:basics}}

\end{figure}

The Simple Boolean Perceptron \cite{Engel01} is defined by a binary
function $\sigma[\boldsymbol{x};\mathbb{\mathcal{\mathsf{\mathfrak{\mathbb{\boldsymbol{J}}}}}}]=sgn\left[\mathbb{J}\cdot\boldsymbol{x}\right]$
that classifies input vectors $\boldsymbol{x}\in\mathbb{R^{N}}$ by
dividing the space into two subspaces, i.e. \textit{for} or \textit{against},
labeled $\left\{ \pm1\right\} $, with the dividing hyperplane specified
by its normal vector $\mathbb{J}\in\mathbb{R^{N}}$ (\emph{internal
representation}) as depicted in Figure \ref{fig:basics}, to the left.
Within the context of opinion dynamics, every agent is represented
by a vector $\mathbb{J}$ that is not accessible to the other agents.
Given a direction in the space of issues, the sign of the overlap
between this and the internal representation determines whether the
agent is favorable or contrary to that issue and the size of this
overlap determines how strongly the agent believes in her judgment.

A Simple Perceptron is capable of learning an unknown rule by processing
a training data set $\mathfrak{D}=\left\{ \left(\boldsymbol{x}^{\mu},S_{\mu}\right)\right\} _{\mu=1}^{P}$,
with $S_{\mu}\in\{\pm1\}$, containing $P$ examples $\left(\boldsymbol{x}^{\mu},S_{\mu}\right)$
of classification by the unknown rule (\emph{training pairs}). The
learning process, however, only converges to the correct classification
function if the unknown rule is linearly separable. The training pairs
can be thought as being generated by another neural network (\emph{teacher})
or by natural phenomena being observed. The learning dynamics consists
in adjusting the internal representation vector $\mathbb{J}$ by minimizing
an error potential $V[\mathbb{J};\mathcal{D}]=\sum_{\mu}V_{\mu}[\mathcal{\mathbb{J}};\boldsymbol{(x}^{\mu},S_{\mu})]$,
that compares the classification rule implemented by $\mathbb{J}$
with the data set $\mathfrak{D}$. This error potential \textit{defines}
a learning algorithm and must be chosen as robustness and optimality
criteria are considered. To give an example, for a Boolean Perceptron
this potential can be simply defined as an error counting function
$V_{\mu}[\mathcal{\mathbb{J}};\boldsymbol{(x}^{\mu},S_{\mu})]=\Theta[-S_{\mu}(\mathbb{J}\cdot\boldsymbol{x}^{\mu})]$,
where $\Theta[x]=1$ if $x\geq0$, and $0$ otherwise, is the Heaviside
step function. In the batch mode the Perceptron extracts the rule
from the entire data set at once, in the on-line mode the Perceptron
is presented to one data point at a time. The on-line dynamics we
study belongs to the scalar modulated hebbian algorithms, given by
\cite{Vicente98} :

\begin{equation}
\mathbb{J}\left(n+1\right)=\mathbb{J}\left(n\right)-\frac{1}{N}\nabla_{\mathcal{\textrm{J}}}V_{\mu}.\label{eq:scalarmodH}\end{equation}
 By defining $\nabla_{\mathcal{\textrm{J}}}V_{\mu}=-\mathcal{W}_{\mu}\boldsymbol{x}^{\mu}S_{\mu}$,
\emph{a modulation} function $\mathcal{W_{\mu}}$, that regulates
the intensity of each modification to the internal representation
vector, and a \emph{hebbian term} $\boldsymbol{x}^{\mu}S_{\mu}$,
that provides the direction for this vector corrections, are introduced
\cite{Engel01}. The scale $\nicefrac{1}{N}$ is chosen such that
sensible macroscopic behavior is produced in the thermodynamic limit
when $\boldsymbol{x}^{\mu}\cdot\boldsymbol{x}^{\mu}=\mathcal{O}(N)$.

Despite its simplicity, Perceptron learning scenarios present a vast
richness of different behaviors, which is derived in part from the
interesting properties of the modulation function $\mathcal{W}$.
For the supervised learning scenario, the modulation function has
been studied exhaustively. One can determine using a variational method
an optimal modulation function $\mathcal{W}^{*}$, in the sense that
leads to the maximal decrease of generalization error \cite{Kinouchi92}
per example. A more fundamental approach in \cite{Opper96} permitted
identifying $\mathcal{W}^{*}$ as a maximum entropy approximation
to on-line Bayesian learning. Evolutionary programming studies \cite{Neirotti03,Caticha06}
showed that, under evolutionary dynamics, learning algorithms evolved
to essentially the optimal algorithms. During evolution a sequence
of dynamical phase transitions signaled the \emph{discovery} of structures
that lead to efficient learning. Interestingly the time order of the
appearance of these structures was stable and robust, occurring in
the same order whenever evolution was successful in obtaining efficient
learning algorithms. This corroborated analytical work about temporal
ordering in \cite{Caticha98}.

Within the class of learning algorithms described by Equation \ref{eq:scalarmodH},
the static learning algorithms, which do not incorporate any annealing
schedule, can be described as learning by the incorporation of two
basic ingredients. These are correlation and error correction. The
pure Hebbian algorithm ($\mathcal{W}=1$) is the classic algorithm
where learning is driven by incorporating correlations between input
and output. The Perceptron algorithm ($\mathcal{W}=1$ if the output
is in error and $0$ if not) is the prime example of learning by correction
of errors. There is a vast literature contrasting learning by error
correction to learning by incorporating correlations in psychology
\cite{Houghton05}. In this paper we consider learning algorithms
which are mixtures of these two strategies, in the context we now
describe.

We here suppose social interactions to be stable and represented by
a directed graph $\mathcal{G}$ with adaptive agents, represented
by Perceptrons, in vertices $a\in V(\mathcal{G})$ (see Figure \ref{fig:basics},
to the right). As agent $a$ interacts, it is influenced by information
that flows from parent $b$ to child $a$ of the arc $(b,a)\in E(\mathcal{G})$
(set of arcs of $\mathcal{G}$). This influence takes place as agent
$a$ learns on line the opinion of $b$, $\sigma_{b}(n)$, on the
public issue at step $n$, $\boldsymbol{x}^{\mu(n)}$:

\begin{equation}
\mathbb{J}_{a}\left(n+1\right)=\mathbb{J}_{a}\left(n\right)+\frac{1}{N}\eta\mathcal{W}_{\delta}\left[h_{a}(n)\sigma_{b}(n)\right]\boldsymbol{x}^{\mu(n)}\sigma_{b}(n),\label{eq:online_learning}\end{equation}

where $\eta$ is the learning rate and $h_{a}(n)=\mathbb{J}_{a}\left(n\right)\cdot\boldsymbol{x}^{\mu}$
is a field that represents the degree of belief on the current opinion
$\sigma_{a}(n)=\mbox{sign}\left[h_{a}(n)\right]$ in a manner akin
to the CODA model \cite{Martins08a}. Notice that in this scenario
a training pair $(\boldsymbol{x}^{\mu(n)},\sigma_{b}(n))$ is composed
by the public issue being discussed and by the opinion of agent $b$
on this issue. Training pairs are provided at each social interaction
with $\sigma_{b}(n)$ playing the role of the classification label
$S_{\mu}$ defined in the general supervised learning scenario described
above. We consider that at each time step $n$ one issue $\mu(n)$
is chosen for debate from the quenched set of $P$ issues $\mathcal{X}=\{\boldsymbol{x}^{\mu}\}_{\mu=1}^{P}$.
Information on revealed opinions can then flow through the arcs of
the graph either synchronously, with all agents accessing neighbor
opinions simultaneously, or asynchronously.

We assume that agents modulate learning according to their agreement,
$h_{a}(n)\sigma_{b}(n)>0$, (or disagreement) before internal representations
are modified. The modulation function is thus a mixture of correlation
seeking and error-correction defined as

\begin{equation}
\mathcal{W}_{\delta}[h_{a}\sigma_{b}]=1-(1-\delta)\Theta\left[h_{a}\sigma_{b}\right],\label{eq:modulation}\end{equation}

where $\delta$ represents how an agent $a$ weights agreement with
its neighbor $b$ in relation to disagreement and $\Theta[x]$ is
the Heaviside step function. 

In the following section we study three simple scenarios on a social
structure defined by a one-dimensional lattice with periodic boundary
conditions (a \emph{ring}), to say: 1. A single public issue with
asynchronous bidirectional information flow; 2. $P$ simultaneous
public issues with asynchronous and bidirectional information flow;
3. $P\rightarrow\infty$ simultaneous issues with synchronous and
unidirectional information flow.

\section{\label{sec:sec3} Patterns of Disagreement on a Ring}

\subsection{\label{sec:sub31} Single issue with asynchronous dynamics}

We start by analyzing the case in which $K$ agents influence each
other asynchronously on a single issue represented by a random vector
$\boldsymbol{x}$ with components drawn \emph{i.i.d.} from a standard
normal distribution such that $\boldsymbol{x}\cdot\boldsymbol{x}=\mathcal{O}(N)$.
At each time step one arc $(b,a)$ is randomly chosen and agent $a$
is influenced by agent $b$. As the issue vector is quenched we can
use Equation \ref{eq:modulation} to rewrite Equation \ref{eq:online_learning}
as the following (deterministic) one-dimensional map:\begin{equation}
h_{a}\left(n+1\right)=h_{a}\left(n\right)+\frac{1}{2}\eta\left\{ \delta\left[\sigma_{b}(n)+\sigma_{a}(n)\right]+\left[\sigma_{b}(n)-\sigma_{a}(n)\right]\right\} \label{eq:1Dmap}\end{equation}

\begin{figure}[h]
\hspace{0.0cm}\includegraphics[scale=0.38]{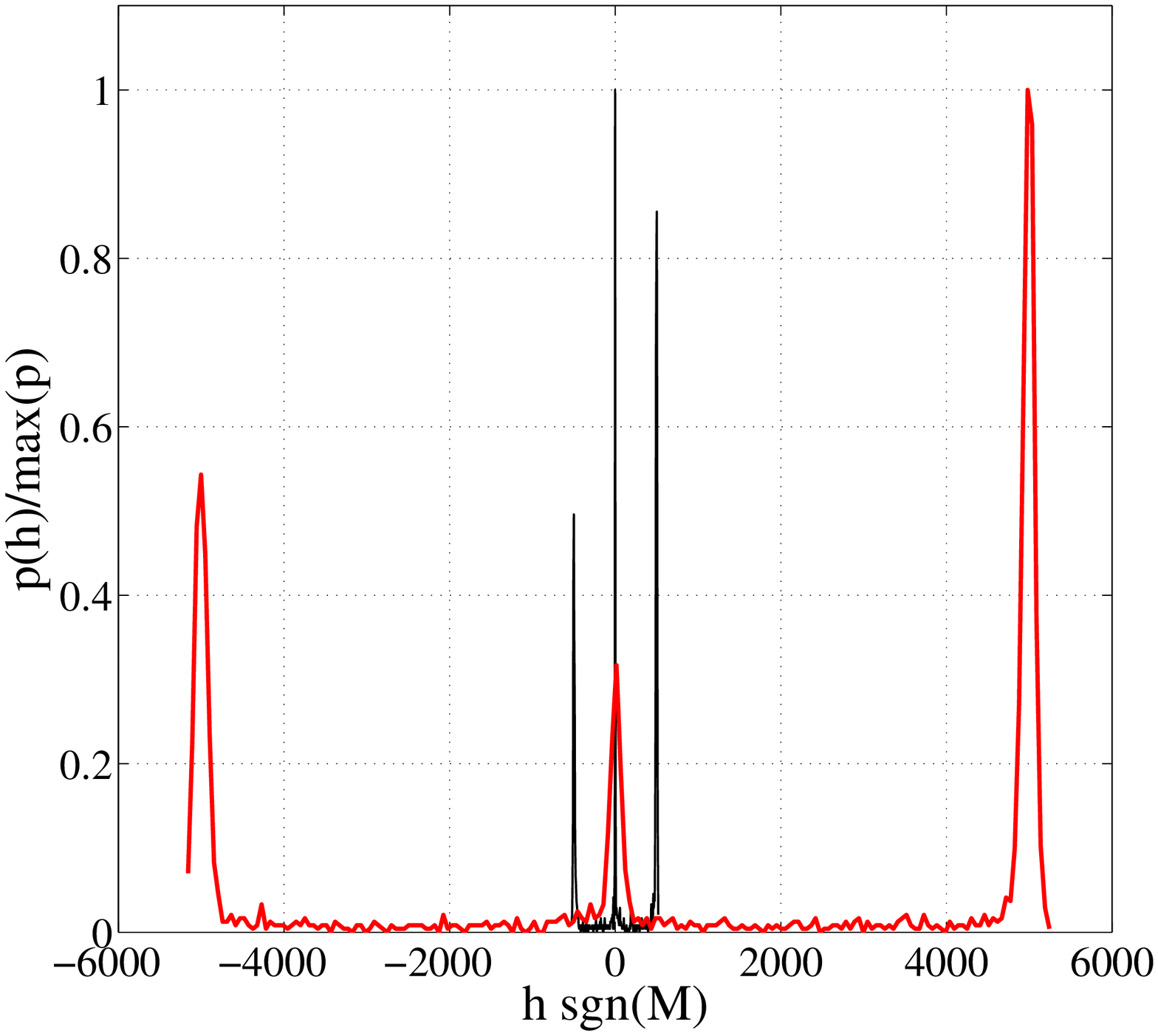}\hspace{-0.4cm}\includegraphics[scale=0.4]{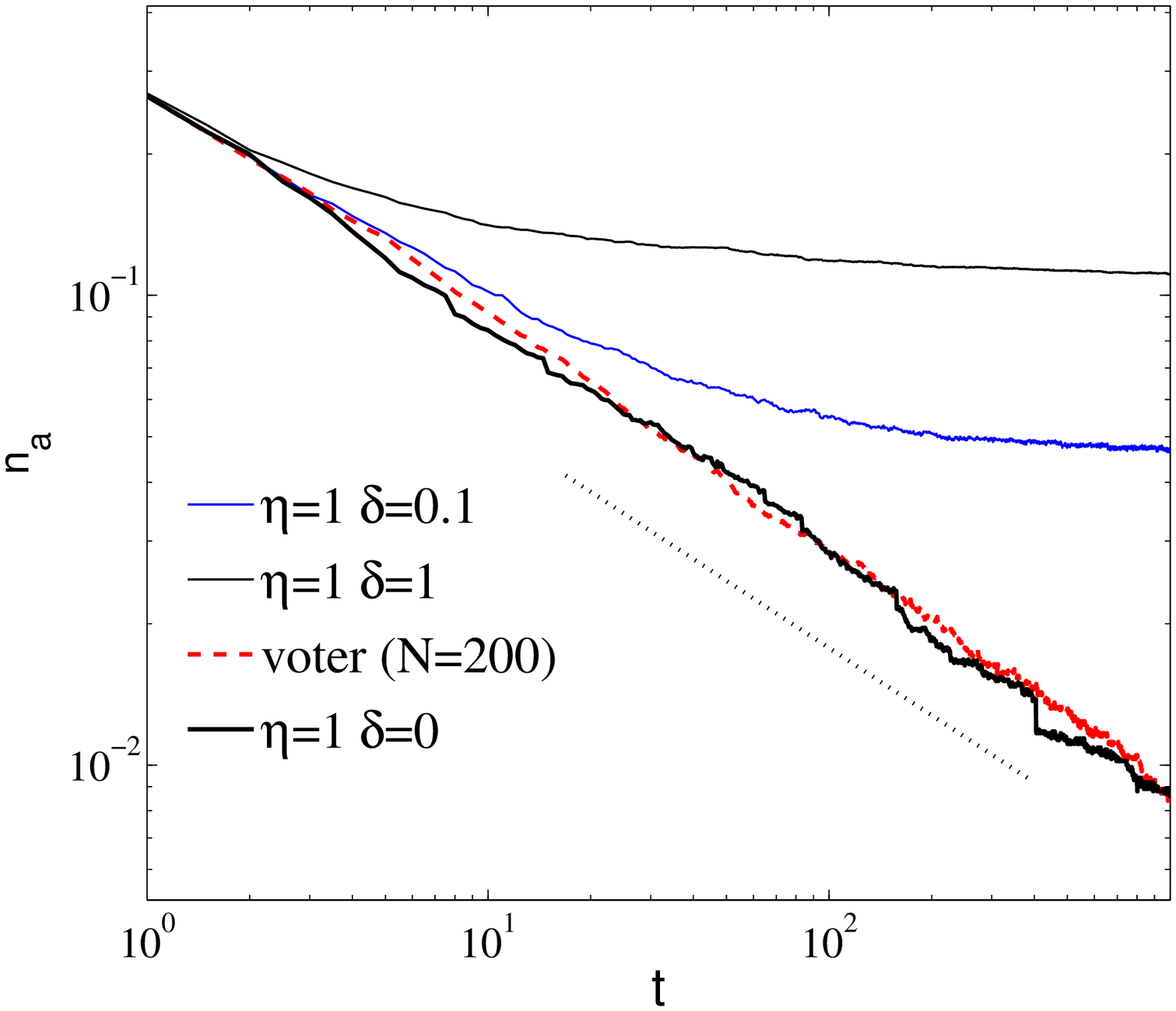}

\caption{Figures represent five simulation runs of an asynchronous dynamics
for $N=200$ and $K=100$ and random initial internal representation
vectors with $\mathbb{J}\cdot\mathbb{J}=1$. \emph{(Left) Distribution
of fields}. Fields are multiplied by the collective opinion and normalized
by the maximum values. The emergence of extremist factions separated
by interfaces, where fields fluctuate around $h=0$, is revealed by
a three peaked distribution. \emph{(Right) Dynamics of the density
of active interfaces}. For $\delta=0$ the system relaxes to consensus
by coarsening, akin to the 1D Voter model, that is also shown. The
dotted line indicates an slope $t^{-\nicefrac{1}{2}}$ expected in
a coarsening process with non-conserved order parameter. For $\delta>0$
the dynamics slows down as fields are mutually reinforced by the neighbors
inside a faction. \label{fig:1} }

\end{figure}

This map is straightforward to simulate in any social graph $\mathcal{G}$,
however, in this paper we will always assume that the social graph
is a ring. For $\delta=1$ agents are in a purely correlation seeking
mode and we recover the CODA model \cite{Martins08a}. For $\delta>0$
the equilibrium state is composed by domains (or extremist factions)
where fields are reinforced by agreement and interfaces that undergo
a random walk resulting (for the one dimensional ring) in a three
peaked distribution $p(h)$ with a peak in $h=0$ produced by the
interfaces and two peaks drifting away symmetrically as $h(n)\thicksim n\eta\delta$.
In Figure \ref{fig:1}, to the left, we show an average over five
field distributions obtained from the simulation of $N=200$, $K=100\;$
system for $\delta=0.1$ and $\delta=1$, with $\eta=1$ in both cases.
Initial internal representation vectors are chosen to be random with
$\mathbb{J}\cdot\mathbb{J}=1$. In order to keep results comparable
among systems of different size $K$, $t$ is measured in MC steps
defined as $t=n/K$. Distributions are normalized by maximum values
and multiplied by the collective opinion given by the sign of $M=\sum_{a=1}^{K}\sigma_{a}$,
to avoid artificially symmetrizing the result upon averaging.

In the case $\delta=0$ agents are purely error-correcting and the
system is driven towards consensus by coarsening (for a social structure
in one dimension) with domains growing with $\sqrt{t}$, akin to Voter
dynamics \cite{Redner01}. In Figure \ref{fig:1}, to the right, we
show the simulated evolution (average over $5$ runs) in time of the
density of active interfaces defined as $n_{a}=(1-\langle\sigma_{a}\sigma_{b}\rangle)/2$,
with $(b,a)\in E(\mathcal{G})$, for $\delta=0$, $\delta=0.1$ and
$\delta=1$ (CODA model). For $\delta>0$ the system is still driven
to order by coarsening, however, the dynamics becomes glassy due to
mutual reinforcing inside factions, leading to $n_{a}(t\rightarrow\infty)>0$
and asymptotically frozen domains.

\subsection{\label{sec:sub32} Set of $P$ issues with asynchronous dynamics}

In the second scenario we analyze, a (quenched) set $\mathcal{X}=\{\boldsymbol{x}^{\mu}\}_{\mu=1}^{P}$
of vectors, with components sampled \emph{i.i.d.} from a standard
normal distribution, represents public issues to be debated. At each
time step $n$ one issue $\boldsymbol{x}^{\mu(n)}\in\mathcal{X}$
and one arc $(b,a)$ are randomly chosen. Agent $a$, then, tries
to learn on-line agent $b$ internal representation $\mathbb{J}_{b}$
employing the learning dynamics described by Equation \ref{eq:online_learning}.

\begin{figure}[h]
\hspace{-0.5cm}\includegraphics[scale=0.4]{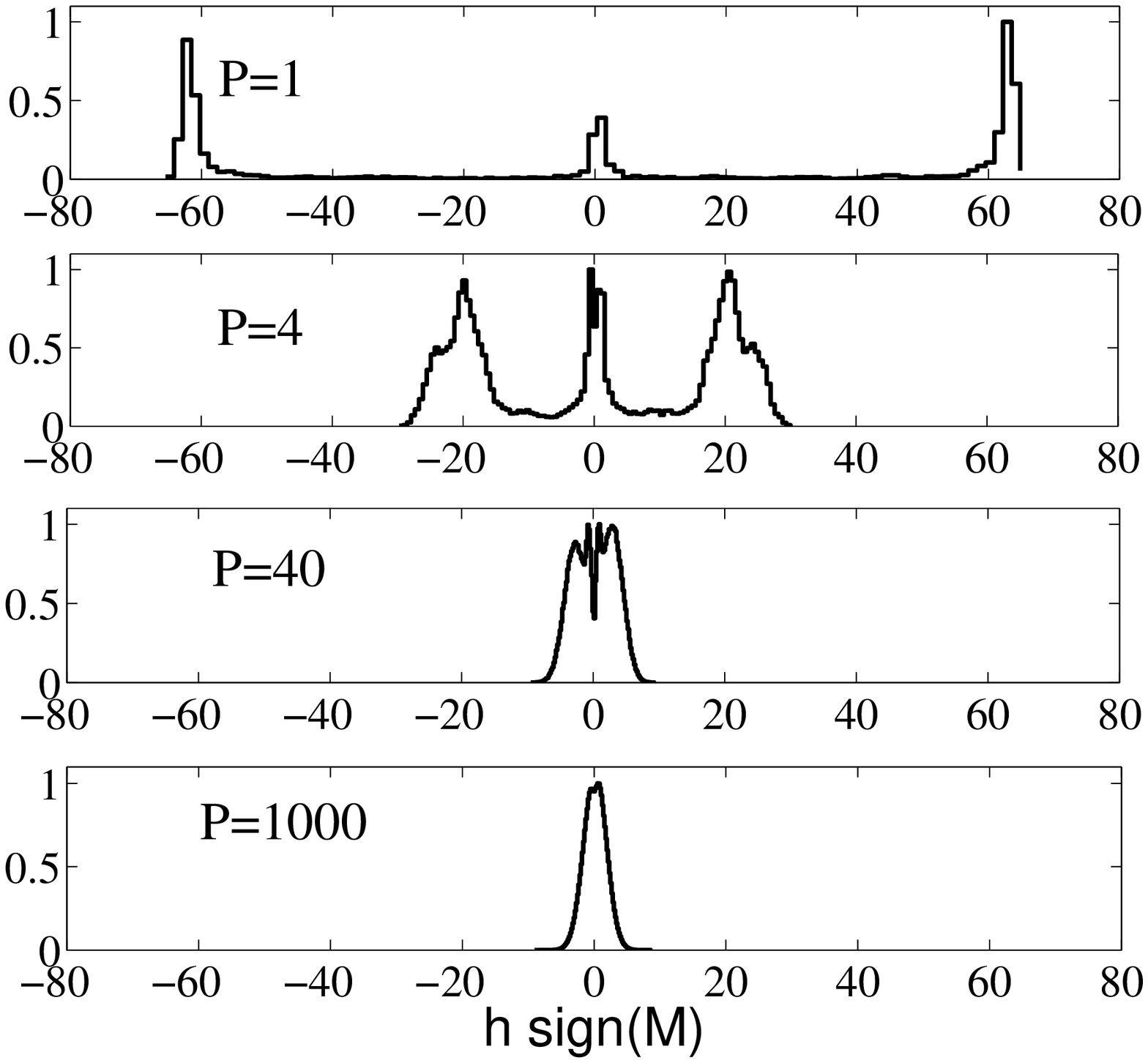}\hspace{-0.5cm}\includegraphics[scale=0.4]{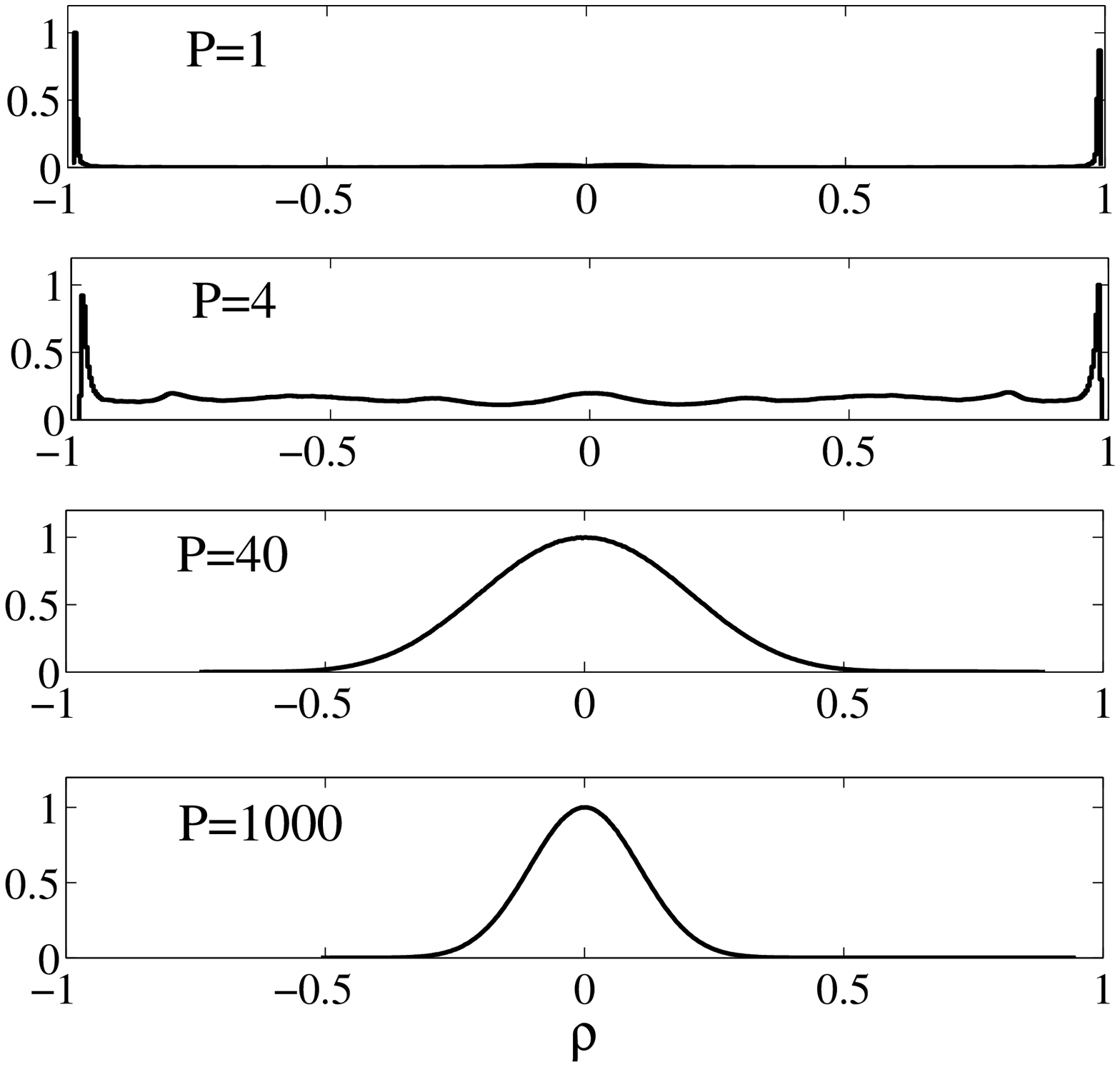}

\caption{Figures show the results of a single run of simulations of asynchronous
dynamics with $P=1,4,40$ and $1000$, $N=100$, $K=10^{4}$, $\eta=1$
and $\delta=0.01$. Measurements are taken for $t=10^{6}$ MC steps
measured as $t=n/K$. \emph{(Left) Distribution of fields with number
of issues}. The histograms show the distribution of belief fields
($h_{a}$). Distributions are normalized by maximum values and multiplied
by the collective opinion, given by the sign of $M=\sum_{a=1}^{K}\sigma_{a}$,
to avoid artificially symmetrizing the result upon averaging. As \emph{$P$}
increases the dynamics slows down and opinions become less extreme
inside factions. \emph{(Right)} \emph{Distribution of overlaps with
number of issues}.\label{fig:Pissues} Opinions become more diverse
as $P$ increases. For $P=1000$ the distribution corresponds to what
is expected to be seen in the case of $K$ random vectors. }

\end{figure}

The restricted quenched set of issues leads to a dynamics that is
not trivially amenable to analytical treatment. In order to get some
insight we have, therefore, relied on a number of simulations on a
1D ring. The general behavior that emerges can be seen on Figure \ref{fig:Pissues}
(Left) that shows, for $P=1,\:4,\:40$ and $1000$, the distribution
of fields multiplied by the collective opinion for a single run with
$N=100$, $K=10^{4}$, $\delta=0.01$ and $\eta=1$. The averages
are taken at time $t=10^{6}$ MC steps measured as $t=n/K$. If agents
are correlation seekers in some measure ($\delta>0$), identifiable
extremist factions emerge for a number $P$ of simultaneous issues
that is not too large. However, as $P$ increases, extremism via field
reinforcement inside a faction weakens. Given that the set of issues
is produced randomly and that $P\ll N$, the probability that two
agents will disagree on a given issue is well defined by $\mathbb{P}\left\{ a\mbox{ disagrees with }b\right\} =\frac{1}{\pi}\arccos\left[\rho_{ab}\right]$,
where $\rho_{ab}=\frac{\mathbb{J}_{a}\cdot\mathbb{J}_{b}}{J_{a\:}J_{b}}$
is the overlap between $a$ and $b$; $J_{a}$ and $J_{b}$ being
vector norms. In Figure \ref{fig:Pissues}, to the right, we show
histograms for overlaps under the same conditions described in the
left panel of the same figure. For $P$ not too large agents either
agree or disagree with certainty, what produces strong mutual reinforcement
effects within a faction. As $P$ increases internal representations
diversify and the probability of disagreement increases reducing extremist
trends. For $P$ very large no convergence of opinions is observed
and the distribution of overlaps is the same expected for random vectors
with $N$dimensions (\emph{i.e.} a dispersion that is $\mathcal{O}\left(\nicefrac{1}{\sqrt{N}}\right)$
is observed).

\begin{figure}[h]
\hspace{2cm}\includegraphics[scale=0.35]{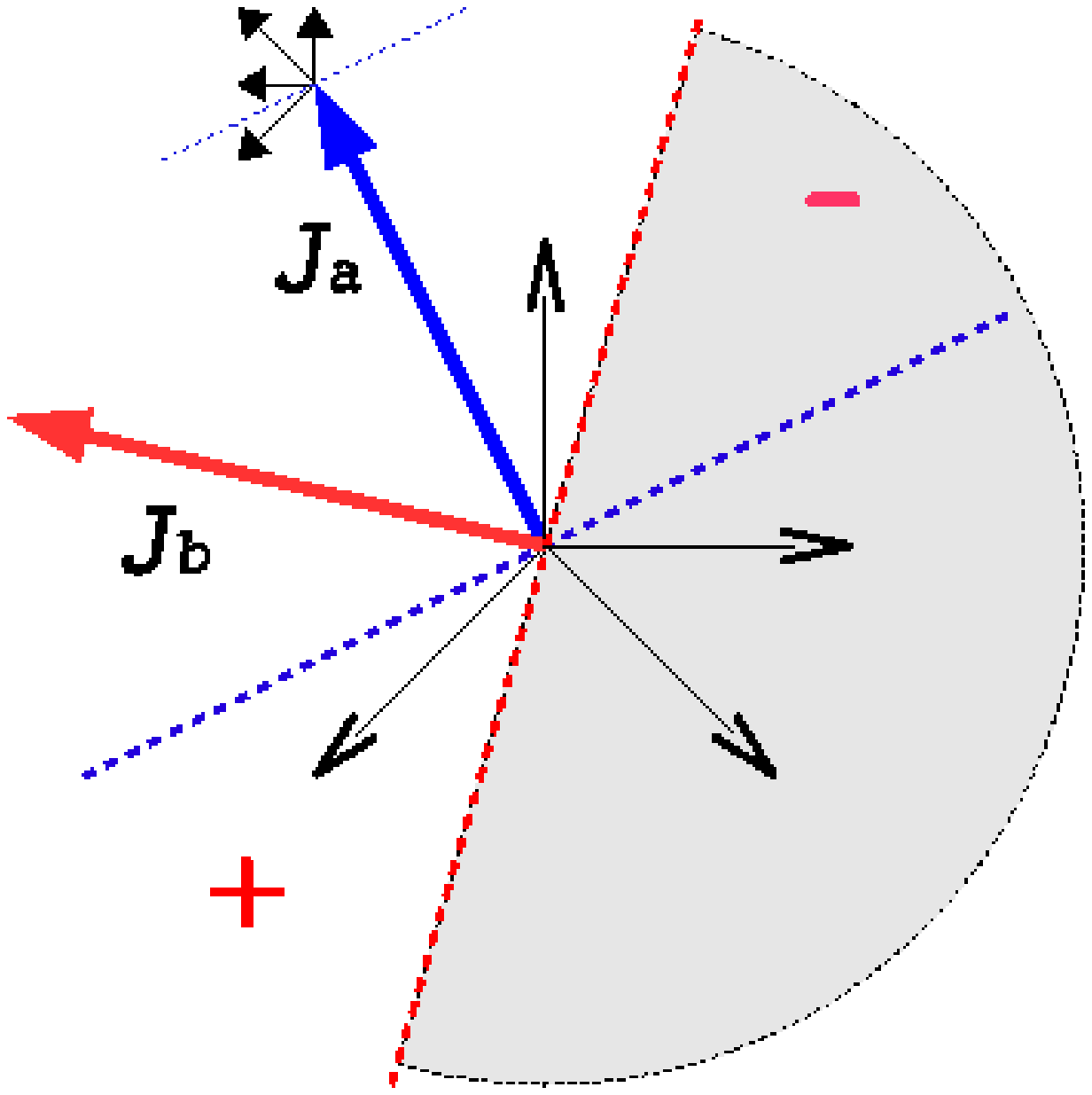}\hspace{0.2cm}\includegraphics[scale=0.4]{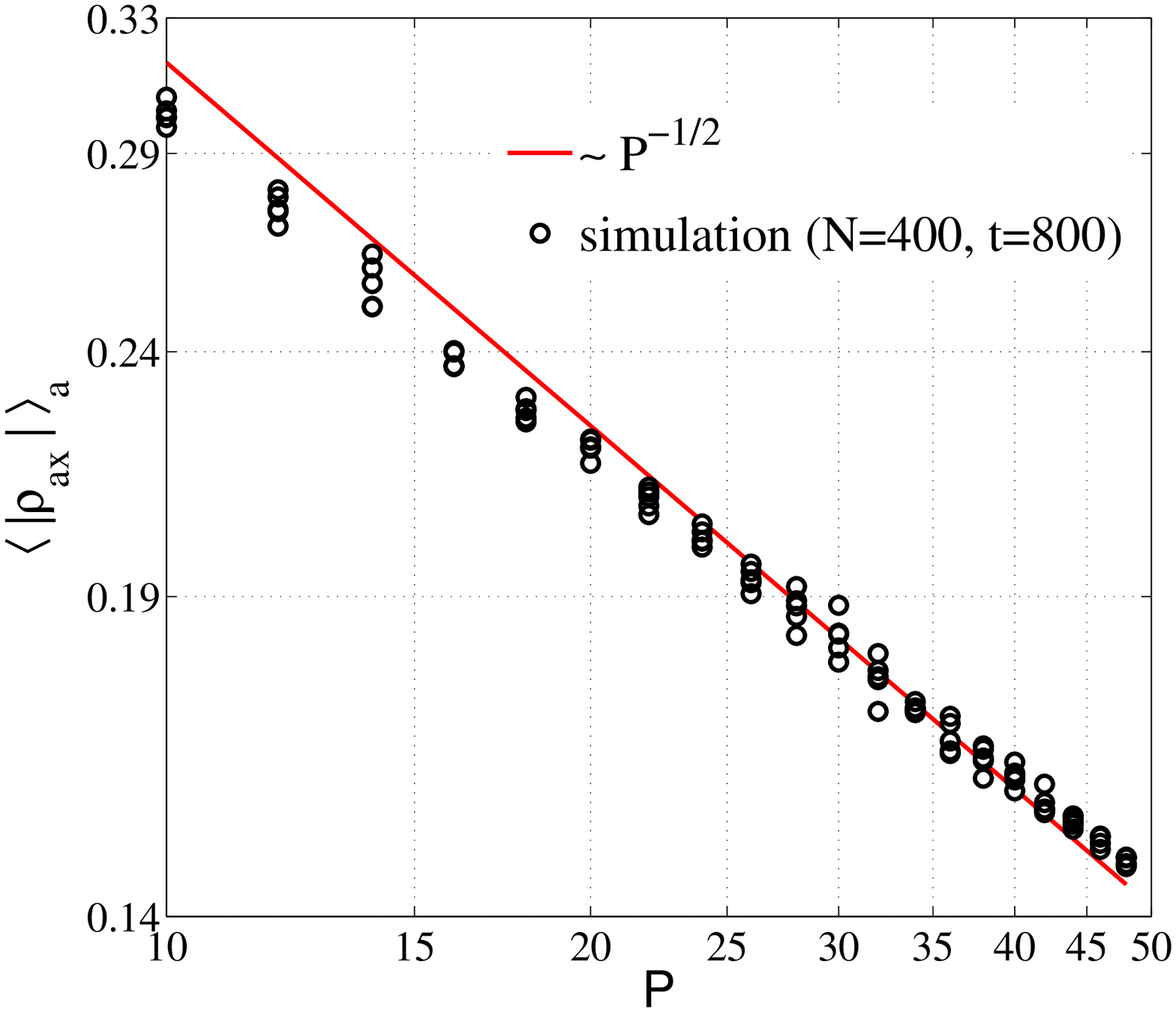}

\caption{\emph{(Left}) \emph{Mutual learning representation.} Inside a faction
overlaps are positive and corrections to the internal representation
$\mathbf{J}_{a}$ due to learning are parallel or anti-parallel to
issues depending on the opinion of agent $b$ (represented by $\mathbf{J}_{b}$).
In the figure the shaded area indicates the half-space where $\sigma_{b}=-1$.
Overlaps between agents only change during learning due to issue components
that are orthogonal to the internal representation being influenced.
The hyperplane orthogonal to $\mathbf{J}_{a}$ is represented by a
thin dashed line in the figure. In the thermodynamic limit $N\rightarrow\infty$
the average displacement due to a sequence of $P$ random issues scales
with $\nicefrac{1}{\sqrt{P}}$. \emph{Asynchronous dynamics.} \emph{(Right)}
\emph{Average overlap to the mean issue as a function of the number
of issues.} Five simulations (circles) with $N=400$, $K=100$, $\eta=1$
and $\delta=1$ are compared to the theoretical prediction for the
scaling with the number of issues $P$. Opinions become more diverse
as the number of simultaneous issues being debated increases. \label{fig:perceptronlearn}}

\end{figure}

No phase transition with $P$ as control parameter takes place though.
Instead, the dynamics slows down continuously with $\nicefrac{1}{\sqrt{P}}$.
A simple argument shows that it is so. Inside a faction two agents
agree in most of the issues, which implies a positive overlap with
neighbors, as depicted in Figure \ref{fig:perceptronlearn}, to the
left. As agent $a$ is influenced by agent $b$, it corrects its internal
representation following the learning dynamics of Equation \ref{eq:online_learning},
which implies that $\mathbb{J}_{a}$ is corrected by $\frac{1}{N}\eta\sigma_{b}\boldsymbol{x}^{\mu(n)}$,
to say, by a vector that is parallel or anti-parallel to the issue
depending on the opinion of agent $b$ being $\sigma_{b}=1$ or $\sigma_{b}=-1$,
respectively. The overlap changes by the average displacement due
to the components of $\frac{1}{N}\eta\sigma_{b}\boldsymbol{x}^{\mu(n)}$
orthogonal to the direction defined by the internal representation
$\mathbb{J}_{a}$. Considering that the dimensionality of the issue
space $N$ is very large ($N\rightarrow\infty$ in the thermodynamic
limit), the $P$ random issues can be considered to be orthogonal
to order $\nicefrac{1}{\sqrt{N}}$. The average displacement due to
a sequence of random issues, therefore, scales with $\nicefrac{1}{\sqrt{P}}$. 

The collective result of the mutual learning of opinions is the alignment
inside a faction of internal representations to a direction defined
by the average issue $\overline{\boldsymbol{x}}=\frac{1}{P}\sum_{\mu=1}^{P}\boldsymbol{x}^{\mu}$.
The typical alignment of agent $a$ internal representation and the
average issue is $\rho_{ax}=\frac{\mathbb{J}_{a}\cdot\boldsymbol{\overline{x}}}{J_{a}\overline{x}}$,
with the mutual learning dynamics being symmetrical in respect to
the hyperplane defined by the average issue. This alignment as a function
of $P$ can be estimated as follows. The norm of the internal representation
grows as $J_{a}\thicksim\sqrt{\frac{n}{N}}$ and the norm of the average
issue scales as $\overline{x}\thicksim\sqrt{\frac{N}{P}}$. On the
other hand, the scaling of the overlap $\mathbb{J}_{a}\cdot\boldsymbol{\overline{x}}$
is estimated as

\begin{eqnarray}
\mathbb{J}_{a}\cdot\left(\frac{1}{P}\sum_{\mu}\boldsymbol{x}^{\mu}\right) & = & \frac{1}{N}\sum_{j=1}^{n}\eta\sigma_{b}(j)\boldsymbol{x}^{\mu(j)}\cdot\left(\frac{1}{P}\sum_{\mu}\boldsymbol{x}^{\mu}\right)\nonumber \\
 & = & \frac{1}{NP}\sum_{j=1}^{n}\eta\sigma_{b}(j)\left(\boldsymbol{x}^{\mu(j)}\cdot\sum_{\mu}\boldsymbol{x}^{\mu}\right)\nonumber \\
 & \sim & \frac{1}{P}\sum_{j=1}^{n}\sigma_{b}(j)\nonumber \\
 & \sim & \frac{\sqrt{n}}{P}\end{eqnarray}

Putting everything together we can finally see that $\rho_{ax}\sim\frac{1}{\sqrt{P}}$.
In the right panel of Figure \ref{fig:perceptronlearn}, we plot the
result of five simulations for $\left|\rho_{ax}\right|$ averaged
over the vertices, where we have dropped the sign to deal with the
learning dynamics symmetry that allows for two factions with $\pm\rho_{ax}$.
In the same figure we also plot a curve $\sim\frac{1}{\sqrt{P}}$
for comparison. Tough not precise, the general agreement is reasonable
considering that we have ignored non-trivial correlations and finite
size effects in the argument above. 

Following along the same line as above it is also interesting to notice
that the dispersion of fields $h_{a}(n)=\mathbb{J}_{a}\cdot\boldsymbol{x}^{\mu(n)}$
depicted in Figure \ref{fig:Pissues}, to the left, scale as $\sigma_{h}\sim\sqrt{\frac{n}{P}}$.
It is, therefore, possible to rescale the dispersion of fields by
redefining time as $n_{P}=\frac{n}{P}$. Both the mean field and the
mean overlap, however, still decay with $\frac{1}{\sqrt{P}}$. 

Considering that time is not rescaled with the number of issues, namely,
that the time allowed for discussion is \emph{not} a function of the
number of issues being discussed we expect that, inside a faction,
agent beliefs become less extreme. The general message is also independent
of the exact decay of $\left|\rho_{ax}\right|$. As the number of
issues being simultaneously debated increases, we expect that inside
a faction agent beliefs become not only less extreme but also more
diverse.

\subsection{\label{sec:sub33} Synchronous dynamics with random issues}

In the third scenario we propose the social graph is again a one dimensional
lattice with arcs of the form $E(\mathcal{G})=\{(a,\, a+1\mbox{(mod \mbox{\mbox{K}})}),\: a=1,...,K\}$
and information flows uni directionally between nodes with internal
representations being updated at once in the whole graph as public
random issues are presented. In this case, at least in principle,
semi-analytical solutions are available for $P$ finite along the
lines of \cite{Coolen00}. For $P\rightarrow\infty$ the scenario
is amenable to semi-analytical treatment by a direct adaptation of
\cite{Metzler00,Engel01,Vicente98} that we summarize in this section.

The technique relies upon the assumption that the macroscopic properties
of the system, defined by the overlaps $R_{ab}=\mathbb{J}_{a}\cdot\mathbb{J}_{b}$,
are self-averaging in the thermodynamic limit ($N\rightarrow\infty$)
\cite{Reents00} and an integration over the distribution of $P\rightarrow\infty$
random issues yields a system of deterministic differential equations.
By supposing that at each time step $n$ an issue $\boldsymbol{x}^{n}$
is produced with components sampled \emph{i.i.d.} from a standard
normal distribution we can write from Equation \ref{eq:online_learning}
difference equations for the overlaps. By scaling time as $\alpha=\frac{n}{N}$
and averaging over the issues we can write a system of $\frac{1}{2}\left(K^{2}+K\right)$
differential equations whose integrals should describe the macroscopic
evolution of the system, given that the self-averaging property holds.
Clearly, the same procedure can be implemented to describe synchronous
dynamics on arbitrary graphs $\mathcal{G}$. On a ring with unidirectional
information flow the first $\frac{1}{2}K(K-1)$ equations describing
internal representation overlaps are given by:

\begin{eqnarray}
\frac{dR_{ab}}{d\alpha} & = & \frac{\eta}{2}\left(\delta+1\right)\left[\langle h_{a}\sigma_{b+1}\rangle+\langle h_{b}\sigma_{a+1}\rangle\right]+\frac{\eta}{2}\left(\delta-1\right)\left[\langle h_{b}\sigma_{a}\rangle+\langle h_{a}\sigma_{b}\rangle\right]\nonumber \\
 & + & \frac{\eta^{2}}{4}\left(\delta-1\right)^{2}\langle\sigma_{a}\sigma_{b}\rangle+\frac{\eta^{2}}{4}\left(\delta+1\right)^{2}\langle\sigma_{a+1}\sigma_{b+1}\rangle\nonumber \\
 & + & \frac{\eta^{2}}{4}\left(\delta+1\right)\left(\delta-1\right)\left[\langle\sigma_{a}\sigma_{b+1}\rangle+\langle\sigma_{b}\sigma_{a+1}\rangle\right],\label{eq:overlap}\end{eqnarray}

where $\langle...\rangle$ denotes averages over the distribution
of issues $\boldsymbol{x}$ that, for our choice, can be calculated
analytically to give

\begin{eqnarray}
\langle\sigma_{a}\sigma_{b}\rangle & = & 1-\frac{2}{\pi}\arccos\left(\frac{R_{ab}}{J_{a}\, J_{b}}\right)\nonumber \\
\langle h_{a}\sigma_{b}\rangle & = & \sqrt{\frac{2}{\pi}}\frac{R_{ab}}{J_{b}},\label{eq:averages}\end{eqnarray}

with $J_{a}=\sqrt{\mathbb{J}_{a}\cdot\mathbb{J}_{a}}$ . The remaining
$K$ equations describe the evolution of the norms of internal representations,
yielding

\begin{eqnarray}
\frac{dJ_{a}}{d\alpha} & = & \frac{\eta}{\sqrt{2\pi}}\left(\delta-1\right)+\frac{\eta^{2}}{2J_{a}}\delta^{2}+\frac{\eta}{\sqrt{2\pi}}\frac{\left(\delta+1\right)}{2}\left[\frac{R_{(a-1)a}}{J_{a-1}\, J_{a}}+\frac{R_{a(a+1)}}{J_{a}\, J_{a+1}}\right]\nonumber \\
 & - & \frac{\eta^{2}}{4\pi J_{a}}\left(\delta^{2}-1\right)\left[\arccos\left(\frac{R_{(a-1)a}}{J_{a-1}\, J_{a}}\right)+\arccos\left(\frac{R_{a(a+1)}}{J_{a}\, J_{a+1}}\right)\right]\label{eq:norms}\end{eqnarray}

when the averages in Equation \ref{eq:averages} are appropriately
replaced.

Along the lines of \cite{Metzler00}, the fixed point of Equation
\ref{eq:norms} can be found for the case of agents that are pure
error-correctors ($\delta=0$). By observing that in this case a synchronous
dynamics preserves the center of mass $\mathbb{J}_{cm}=\frac{1}{K}\sum_{a=1}^{K}\mathbb{J}_{a}$
and by supposing a symmetric evolution, with $R=R_{ab}$ and $J=J_{a}$
for all $a$ and $b$, we can rewrite Equation \ref{eq:norms} as

\begin{equation}
\frac{dJ}{d\alpha}=-\frac{\eta}{\sqrt{2\pi}}+\frac{\eta}{\sqrt{2\pi}}\frac{R}{J^{2}}+\frac{\eta^{2}}{2\pi J}\arccos\left(\frac{R}{J^{2}}\right)\label{eq:norm_symmetric}\end{equation}

which has a fixed point given by

\begin{equation}
\frac{\eta}{J_{cm}}=\frac{\sqrt{2\pi}}{\arccos\left(\rho\right)}\frac{1-\rho}{\sqrt{1+\left(K-1\right)\rho}},\label{eq:fixedpoint}\end{equation}

where $\rho=\nicefrac{R}{J^{2}}$. In the left panel of Figure \ref{fig:FIg5},
we show the average overlap as a function of the normalized learning
rate $\nicefrac{\eta}{J_{cm}}$. Ten simulations (symbols) with $N=100$,
$K=100$ (circles) and $K=1000$ (squares) at $\alpha=200$ are shown
and compared to fixed points predicted by Equation \ref{eq:fixedpoint}.
Interesting features of the synchronous case are that consensus of
error-correcting agents is fostered by slow learning ($\eta\rightarrow0$),
what is also observed in \cite{Stark08}, and that a system of adaptive
agents under this condition tend towards a random distribution of
opinions ($\rho\rightarrow0$) as the system grows larger.

\begin{center}
\begin{figure}[h]
 \hspace{-0.1cm}\includegraphics[clip,scale=0.4]{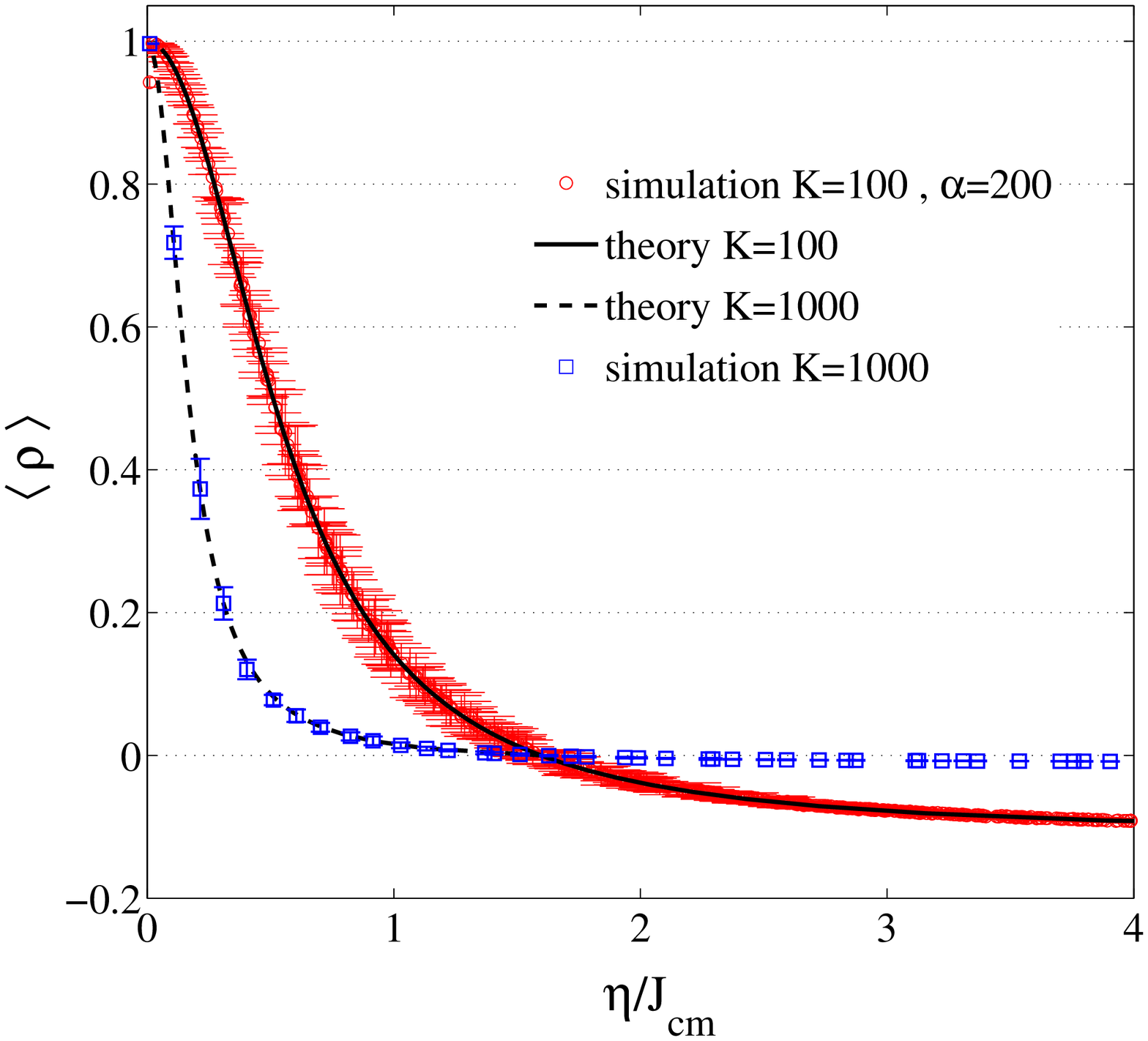}\hspace{-0.6cm}\includegraphics[scale=0.4]{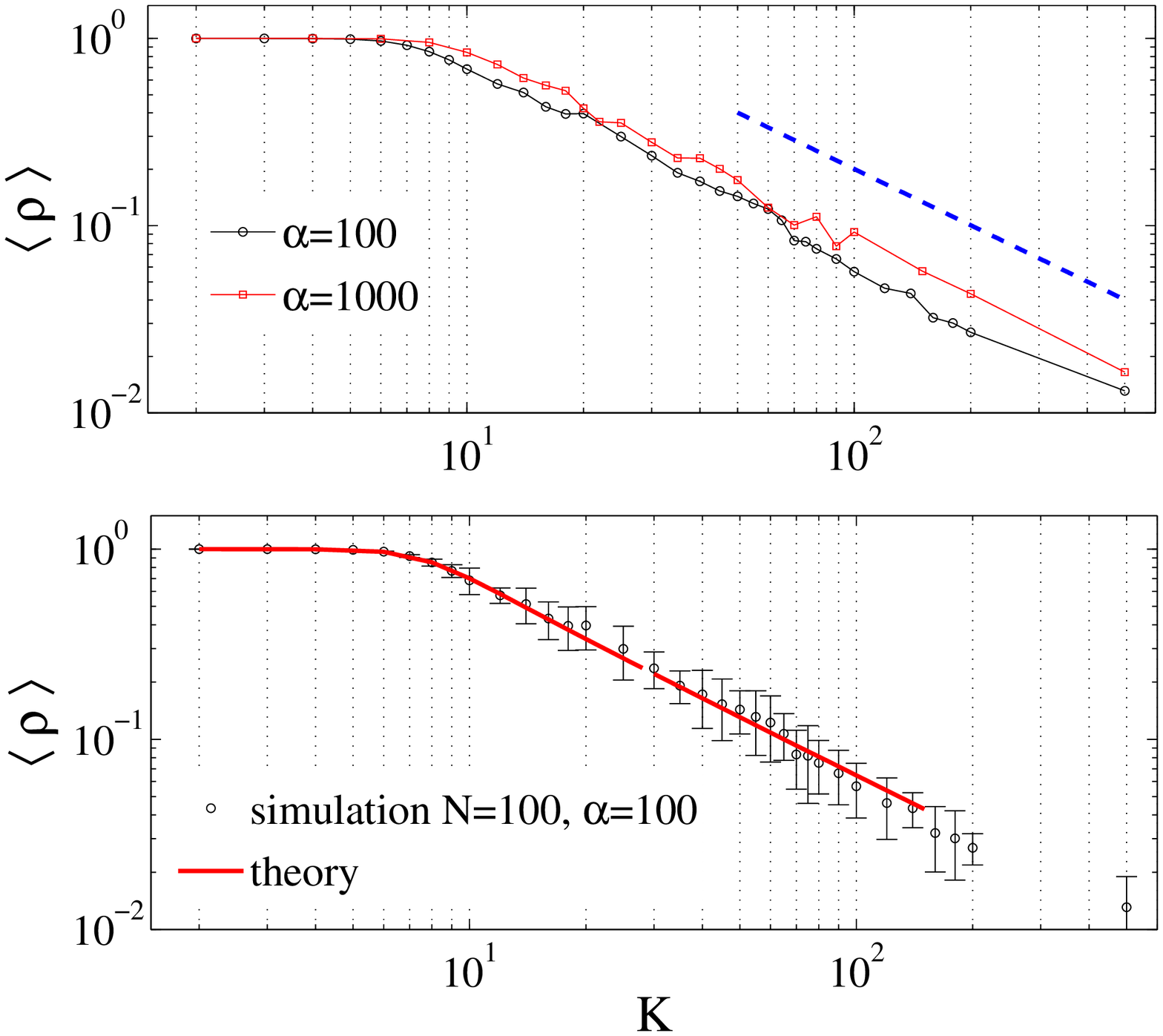}

\caption{\emph{Synchronous Dynamics. (Left) Average overlap versus learning
rate for error-correctors $(\delta=0)$.} Symbols show overlaps averaged
over arcs $(a,b)\in E(\mathcal{G})$ in ten simulations with $N=100$,
$K=100$ (circles) and $K=1000$ (squares) at $\alpha=200$. Lines
represent the fixed points given by Equation \ref{eq:fixedpoint}.
Consensus is favored by small learning rates ($\eta\rightarrow0$).
\emph{(Right) Average overlap versus system size for correlation seekers
$(\delta=1)$.} Top: Averages are over vertices and 10 simulation
runs with $N=100$, $\eta=1$, $\delta=1$ at, from bottom up, $\alpha=100$
(black) and $\alpha=1000$(red). The dashed line indicates the slope
of $K^{-1}$. Bottom: Numerical integration of Equations \ref{eq:overlap}
and \ref{eq:norms} compared to simulations at $\alpha=100$. Consensus
is attained even for $P\rightarrow\infty$, given that the system
size is not too large. For large $K$ the asymptotic average overlap
scales as $\langle\rho\rangle\sim\nicefrac{1}{K}$. \label{fig:FIg5}}

\end{figure}

\par\end{center}

For $\delta>0$ internal representations are constantly reinforced
by confirmation inside a faction and no fixed point can be found for
Equation \ref{eq:norms}. To simplify the analysis we assume the extreme
case of agents that are pure correlation seekers $\delta=1$. Introducing
$\lambda_{a}=\nicefrac{\eta}{J_{a}}$ and $\rho_{ab}=\nicefrac{R_{ab}}{J_{a}J_{b}}$
and considering the limit of long times ($\lambda_{a}$ small) leads
to:

\begin{equation}
\frac{d\rho_{ab}}{d\alpha}\approx\sqrt{\frac{2}{\pi}}\lambda_{a}\left(\rho_{(a+1)b}-\rho_{ab}\rho_{a(a+1)}\right)+\sqrt{\frac{2}{\pi}}\lambda_{b}\left(\rho_{a(b+1)}-\rho_{ab}\rho_{b(b+1)}\right)\label{eq:overlap_longtimes}\end{equation}

and

\begin{equation}
\frac{d\lambda_{a}}{d\alpha}\approx-\sqrt{\frac{2}{\pi}}\lambda_{a}^{2}\rho_{a(a+1)}.\label{eq:norm_longtimes}\end{equation}

These equations imply that the state of consensus (\emph{$\rho_{ab}=1$}
for all \emph{$a$} and $b$) is always a fixed point, however, the
dynamics slows down with $\lambda_{a}$ as the system drifts towards
the fixed point defined by an arbitrary distribution of overlaps and
$\lambda_{a}=0$.

In Figure \ref{fig:FIg5}, to the top right, we show the average overlaps
(over vertices and over ten runs) as a function of the system size
$K$ for simulations with $N=100$, $\eta=1$, $\delta=1$ at, from
bottom up, $\alpha=100$ and $\alpha=1000$. In Figure \ref{fig:FIg5},
to the bottom right, we show a comparison between the result of simulations
at $\alpha=100$ and the numerical solution for Equations \ref{eq:overlap}
and \ref{eq:norms}. In contrast with the asynchronous case, if the
update is synchronous consensus can be attained even for $P\rightarrow\infty$,
given that the system size $K$ is not too large. For large $K$ the
asymptotic average overlap decays as $\langle\rho\rangle\sim\nicefrac{1}{K}$
(dashed line to the top right of Figure \ref{fig:FIg5}).

\section{Conclusions}

A central puzzle in the study of information processing in social
systems is the persistence of disagreement in a system composed by
consensus seekers. This puzzle can be solved by considering a dichotomy
between internal representations and revealed opinions, as the introduction
of memory effects may allow for competition between a local trend
toward uniform opinions and long range heterogeneity. On top of that,
in the model for opinion dynamics with learning we have introduced,
agreement can be attained in one issue while disagreement persists
in other issues.

In the model we have proposed agents classify any issue presented
to discussion by using a single internal representation $\mathbb{J}$
which implies the strong presupposition that agents are always consistent
when issuing opinions. It seems more realistic also considering cases
with multiple internal representations or with noisy decision processes
allowing for inconsistent judgments. Such directions, however, still
have unclear consequences and might be pursued in further work. 

We have shown that some simple instances of the model are amenable
to analytical treatment with a number of nontrivial phenomena emerging.
For a single issue with asynchronous update of information and a one
dimensional social structure, we observe behavior akin to the Voter
model as long as agents are pure error-correctors ($\delta=0$), with
full consensus being reached by a coarsening process. When agents
are also correlation seekers ($\delta\neq0$), extremist factions
with opposite opinions survive due to the mutual reinforcement of
confirmed beliefs. That is, extreme disagreement becomes a characteristic
of the system, even in the long run. The survival of disagreement,
however, comes in this case at the cost of the appearance of extremist
agents.

By introducing a number of issues to be simultaneously debated, we
observe that the tendency towards extremism weakens. This seems to
suggest that a richer cultural setting, where the agents debate a
larger number of different subjects is less likely to develop extremists
than a setting where very few issues (or highly correlated issues)
are debated. As the number of issues grows larger, internal representations
converge to distributions that are indistinguishable from random vectors
in a space of equal dimensionality. Disagreement also survives in
this case, but with the tendency to extremism tamed.

We also have investigated semi-analytically the case of synchronous
information update with a very large number of simultaneous issues.
A dynamics distinct from the asynchronous case is observed, with consensus
being fostered either by slowing down the adaptation process, when
agents are error-correctors, or by allowing agents to be correlation
seekers, given that the system is not too large.

\ack This work has been funded by Conselho Nacional de Desenvolvimento
Científico e Tecnológico (CNPq), under grant 550981/2007-1 (RV), and
by Fundação de Amparo à Pesquisa do Estado de São Paulo (FAPESP),
under grant 2008/00383-9 (ACRM). We also wish to thank the anonymous
referee for his thoughtful suggestions which contributed to improve
our manuscript.


\section*{References}

\begin{thebibliography}{10}
\bibitem[1]{Johnson04}Huckfeldt R, Johnson PE and Sprague J. 2004.
\emph{Political Disagreement: The Survival of Diverse Opinions within
Communication Networks} (Cambridge: CUP)

\bibitem[2]{Sunstein05}Sunstein CR. 2005. \emph{Why Societies Need
Dissent} (Cambridge MA: Harvard University Press)

\bibitem[3]{Huckfeldt95}Huckfeldt R et al. 1995. \emph{American Journal
of Political Science}, \textbf{39}, 4, 1025.

\bibitem[4]{Castellano07}Castellano C, Fortunato S, and Loreto V.
2007. \emph{arXiv:physics.soc-ph/0710.3256}

\bibitem[5]{Galam82}Galam S, Gefen Y and Shapir Y. 1982. \textit{J.
Math. Sociol.}, \textbf{9}, 1

\bibitem[6]{Clifford73} Clifford P and Sudbury A. 1973. \textit{Biometrika},
\textbf{60}, 581

\bibitem[7]{Holley75}Holley R and Liggett T. M. 1975. \textit{Ann.
Probab.} \textbf{3}, 643

\bibitem[8]{Sznajd00}Sznajd-Weron K. and Sznajd, J. 2000. \textit{Int.
J. Mod. Phys. C} \textbf{11}, 1157

\bibitem[9]{Latane81}Latané B. 1981. \textit{Am. Psychol.}, \textbf{36},
343

\bibitem[10]{Lewenstein92}Lewenstein M, Nowak A and Latané B. 1992.
\textit{Phys. Rev. A}, \textbf{45}(2), 763

\bibitem[11]{Martins08a} Martins ACR. 2008. \emph{Int. J. Mod. Phys.
C} \textbf{19} 617

\bibitem[12]{Axelrod97}Axelrod R. 1997. \textit{Journal of Conflict
Resolution}, \textbf{41}(2), 203

\bibitem[13]{Castellano00} Castellano C, Marsili M and Vespignani
A. 2000. \emph{Phys. Rev. Lett.}, \textbf{85}, 3536

\bibitem[14]{Engel01} Engel A and Van den Broeck C. 2001. \emph{Statistical
Mechanics of Learning} (Cambridge: CUP)

\bibitem[15]{Stauffer08} Stauffer D, Grabowicz PA and Holyst JA.
2008. \emph{arXiv:physics.soc-ph/0712.4364v2}

\bibitem[16]{Vicente98}Vicente R, Kinouchi O and Caticha N. 1998.
\emph{Machine Learning} \textbf{\emph{32}} 179

\bibitem[17]{Metzler00}Metzler R, Kinzel W and Kanter I. 2000. \emph{Phys.
Rev. E} \textbf{\emph{62}} 2555

\bibitem[18]{Kinouchi92}Kinouchi O and Caticha N. 1992. \emph{J.
Phys. A: Math. and Gen.} \textbf{25} 6243

\bibitem[19]{Opper96}Opper M. 1996. \emph{Phys. Rev. Lett.} \textbf{77}
4671

\bibitem[20]{Neirotti03}Neirotti JP and Caticha N. 2003. \emph{Phys.
Rev. E} \textbf{67} 041912

\bibitem[21]{Caticha06}Caticha N and Neirotti JP. 2006. \emph{The
evolution of learning systems: to Bayes or not to be.} In \emph{Bayesian
Inference and Maximum Entropy Methods in Science and Engineering,
26th International Workshop}, edited by A. Mohammad-Djafari.

\bibitem[22]{Caticha98}Caticha N and Kinouchi O. 1998. \emph{Phil.
Mag B} \textbf{77} 1565

\bibitem[23]{Houghton05}Houghton G (ed.). 2005. \emph{Connectionst
Models in Cognitive Psychology} (Routledge)

\bibitem[24]{Redner01}Redner S. 2001. \emph{A Guide to First-Passage
Processes} (Cambridge:CUP)

\bibitem[25]{Coolen00}Coolen ACC and Saad D. 2000. \emph{Phys. Rev.
E} \textbf{62} 5444

\bibitem[26]{Reents00}Reents G and Urbanczik R. 2000. \emph{Phys.
Rev. Lett.} \textbf{80} 5445

\bibitem[27]{Stark08}Stark H-U, Tessone CJ and Schweitzer F. 2008.
\emph{Phys. Rev. Lett.} \textbf{101} 018701 
\end{thebibliography}
\end{document}